\documentclass[aps,twocolumn,showpacs,superscriptaddress]{revtex4}
\usepackage{epsfig}
\usepackage{times}

\begin{document}

\title{Zealots tame oscillations in the spatial rock-paper-scissors game}

\author{Attila Szolnoki}
\email{szolnoki@mfa.kfki.hu}
\affiliation{Institute of Technical Physics and Materials Science, Centre for Energy Research, Hungarian Academy of Sciences, P.O. Box 49, H-1525 Budapest, Hungary}

\author{Matja{\v z} Perc}
\email{matjaz.perc@uni-mb.si}
\affiliation{Faculty of Natural Sciences and Mathematics, University of Maribor, Koro{\v s}ka cesta 160, SI-2000 Maribor, Slovenia}
\affiliation{CAMTP -- Center for Applied Mathematics and Theoretical Physics, University of Maribor, Krekova 2, SI-2000 Maribor, Slovenia}

\begin{abstract}
The rock-paper-scissors game is a paradigmatic model for biodiversity, with applications ranging from microbial populations to human societies. Research has shown, however, that mobility jeopardizes biodiversity by promoting the formation of spiral waves, especially if there is no conservation law in place for the total number of competing players. Firstly, we show that even if such a conservation law applies, mobility still jeopardizes biodiversity in the spatial rock-paper-scissors game if only a small fraction of links of the square lattice is randomly rewired. Secondly, we show that zealots are very effective in taming the amplitude of oscillations that emerge due to mobility and/or interaction randomness, and this regardless of whether the later is quenched or annealed. While even a tiny fraction of zealots brings significant benefits, at 5\% occupancy zealots practically destroy all oscillations regardless of the intensity of mobility, and regardless of the type and strength of randomness in the interaction structure. Interestingly, by annealed randomness the impact of zealots is qualitatively the same as by mobility, which highlights that fast diffusion does not necessarily destroy the coexistence of species, and that zealotry thus helps to recover the stable mean-field solution. Our results strengthen the important role of zealots in models of cyclic dominance, and they reveal fascinating evolutionary outcomes in structured populations that are a unique consequence of such uncompromising behavior.
\end{abstract}

\maketitle

\section{Introduction}
Despite of its simplicity, the rock-paper-scissors game is popular not just for settling everyday disputes in a quick, luck-dependent manner, but also as the basis for research aimed at explaining the intriguing biodiversity in nature \cite{frachebourg_prl96, frean_prsb01, kerr_n02, kirkup_n04, mobilia_pre06, reichenbach_n07, reichenbach_prl08, mobilia_jtb10, kelsic_n15}. Cyclical interactions are also common in evolutionary games with three or more competing strategies, such as in the public goods game with positive and negative incentives \cite{szolnoki_prx13}, in the ultimatum game with discrete strategies \cite{szolnoki_prl12}, as well as in pairwise social dilemmas with coevolution \cite{szolnoki_epl09} or jokers \cite{requejo_pre12b}. Prominent experimental observations of cyclic dominance include the mating strategy of side-blotched lizards \cite{sinervo_n96}, overgrowth of marine sessile organisms \cite{burrows_mep98}, genetic regulation in the repressilator \cite{elowitz_n00}, and competition in microbial populations \cite{durrett_jtb97, kirkup_n04, neumann_gf_bs10, nahum_pnas11}.

The spatial rock-paper-scissors game, where the interaction range of each individual player is limited to its directly linked neighbors, has a long and fruitful history in statistical physics research \cite{szabo_pre99, frean_prsb01, de-oliveira_prl02, reichenbach_pre06, peltomaki_pre08, peltomaki_pre08b, laird_jtb09, berr_prl09, he_q_pre10, wang_wx_pre10b, ni_x_c10, mathiesen_prl11, avelino_pre12, jiang_ll_pla12, roman_jsm12, szczesny_epl13, avelino_pre12b, juul_pre12, roman_pre13, hua_epl13, knebel_prl13, juul_pre13, avelino_pre14, szczesny_pre14, rulquin_pre14, laird_oikos14, szolnoki_pre14c, bose_ijb15, javarone_epjb16, knebel_nc16, roman_jtb16}, not least because some experiments attest to the fact that spatial structure may be just as important for biodiversity as cyclical interactions themselves. For example, experiments with \textit{Escherichia coli} have revealed that arranging the bacteria on a Petri dish is crucial for keeping all three competing strains alive \cite{kerr_n02, kerr_n06}.

Almost a decade ago, Reichenbach et al.~\cite{reichenbach_n07} have shown that the mobility of players in the rock-paper-scissors promotes the formation of spiral waves, which jeopardizes biodiversity when the wavelength exceeds the linear size of the system. Global, system-wide oscillations of the density of the three strategies thus emerge above a critical threshold value of mobility, and as a result the system can easily terminate into a homogeneous state with only a single strategy present. However, Peltom{\"a}ki and Alava~\cite{peltomaki_pre08} have subsequently shown that the formation of spiral waves does not occur if there is a conservation law in place for the total number of competing players -- a condition which was not met in \cite{reichenbach_n07}. If the total number of players is conserved, then mobility has no particular impact on diversity because oscillations are damped by the conservation law.

Here we extend the scope of the spatial rock-paper-scissors game by considering a setup where the total number of competing players is preserved, but where in addition either quenched or annealed randomness is introduced to the square lattice, and where a fraction of the population is occupied by zealots. Mobilia~\cite{mobilia_prl03} has shown that zealotry can have a significant impact on the segregation in a two-state voter model, which suggests that their presence is likely to be significant also in the rock-paper-scissors game with intransitive relationships (for research considering protection spillovers see \cite{kelsic_n15, szolnoki_njp15}). Zealots are players that never change their strategy, regardless of the neighborhood. Such uncompromising behavior exists in human societies, where we have stubborn voters and staunch proponents of ideologies, but can also be observed in other natural systems, including microbial populations, where a mutation might grant a few selected microbes an evolutionary escape hatch out of the closed loop of dominance.

As we will show in what follows, even if a conservation law for the total number of players applies, mobility still jeopardizes biodiversity if only a small fraction of links of the square lattice is randomly rewired. We will also show that zealots are very effective in taming the amplitude of oscillations that is due to mobility and interaction randomness. Our results corroborate recent research concerning the rock-paper-scissors game in well-mixed populations \cite{verma_pre15}, where it was shown that zealotry promotes coexistence. In large structured populations, however, zealotry leads to further fascinating evolutionary outcomes that are a unique consequence of this uncompromising behavior. Before going into details, we first present the definition of the spatial rock-paper-scissors game with mobility and zealots, and the details of the Monte Carlo simulation procedure.

\section{Spatial rock-paper-scissors game with mobility and zealots}

As the basis, we consider the classical rock-paper-scissors game, where the three species cyclically dominate each other. For convenience, we refer to the species as $R$, $P$ and $S$, where strategy $R$ invades strategy $S$, strategy $S$ invades strategy $P$, and strategy $P$ invades strategy $R$. To extend this basic setup, we assume that a fraction $\mu$ of players are zealots, who never change their strategy during the evolution, and this independently of their neighbors. To avoid any bias, we assume that all three possible strategies are initially equally represented among zealots.

The game is studied in a structured population, such that each player is located on the site $x$ of a square lattice with periodic boundary conditions, where the grid contains $L \times L$ sites. In addition, we explore the impact of interaction randomness \cite{watts_dj_n98}, which has proven to be a decisive factor before \cite{szabo_jpa04, szolnoki_pre04b}. We consider the impact of quenched and annealed randomness separately. Quenched randomness is introduced by randomly rewiring a fraction $Q$ of the links that form the square lattice whilst preserving the degree $z=4$ of each site. We thereby obtain regular small-world networks for small values of $Q$ and a regular random network in the $Q \to 1$ limit. Importantly, the rewiring is performed only once before the start of the game, thus introducing quenched (time invariant) randomness in the interactions among the players. In the alternative version of our model annealed randomness is introduced so that at each instance of the game a potential target for an invasion is selected randomly from the whole population with probability $P$, while with probability $1-P$ the invasion is restricted to a randomly selected nearest neighbor \cite{szabo_jpa04, szolnoki_pre04b}. For $P=1$ we thus obtain well-mixed conditions, while for $P=0$ only short-range invasions along the original square lattice interaction structure are possible.

The evolution of strategies proceeds in agreement with a random sequential update, where during a full Monte Carlo step (MCS) every player receives a chance once on average to invade one randomly selected neighbor (or any member of the population with probability $P$ in case of annealed randomness). To introduce mobility, during an elementary step we choose a nearest-neighbor pair randomly where players exchange their positions with probability $\sigma$. In the alternative case, which happens with probability $1-\sigma$, the dominant strategy invades the other position in agreement with the rules of the rock-paper-scissors game. In this way $\sigma$ characterizes the intensity of mobility, whilst ensuring that the number of players is conserved (which prohibits the emergence of spiral waves and oscillations on a regular lattice \cite{peltomaki_pre08}).

During the evolutionary process, we monitor the concentration of each strategy, and we characterize global oscillations with the order parameter $A$, which we define as the area of the limit cycle in the ternary diagram \cite{szabo_jpa04}. This order parameter is zero when the system is in the $\rho_R=\rho_P=\rho_S=1/3$ stationary state and becomes one when the system terminates into an absorbing, one-strategy state. We have used lattices with up to $L \times L = 4 \cdot 10^6$ sites, which was large enough to avoid accidental fixations when the amplitude of oscillations was large, and which allowed an accurate determination of strategy concentrations that are valid in the large size limit. Naturally, the relaxation time depends sensitively on the model parameters and the system size, but $5 \cdot 10^5$ MCS was long enough even for the slowest evolution that we have encountered during this study.

\section{Results}

\begin{figure}
\centerline{\epsfig{file=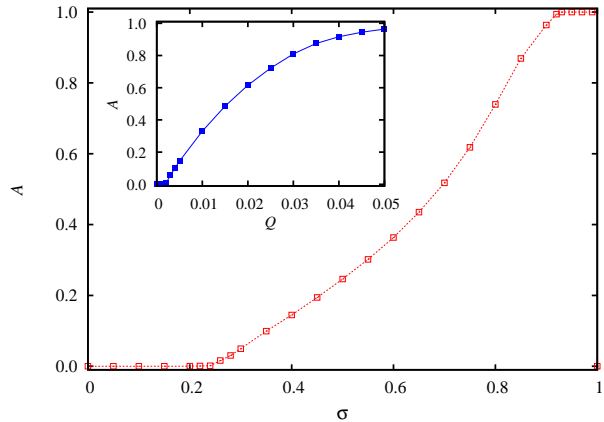,width=8.5cm}}
\caption{(Color online) Oscillations emerge at a critical intensity of mobility if the interaction lattice contains at least some randomness. The main panel shows the area of the limit cycle in the ternary diagram $A$ in dependence on the intensity of mobility $\sigma$, as obtained for the fraction of rewired links $Q=0.05$. The inset shows $A$ in dependence on $Q$, as obtained for $\sigma=0.9$. Evidently, even a minute fraction of rewired links ($Q=0.002$) suffices to evoke non-zero $A$ values if the mobility is high. These results were obtained without zealots ($\mu=0$).}
\label{random}
\end{figure}

Our first result is that departing from the regular square lattice interaction network by introducing some fraction $Q$ of rewired links drastically changes the pattern formation in the face of mobility. In particular, even if the mobility is strong, on the square lattice spiral waves never emerge if the total number of players is conserved (as is presently the case). But if some randomness is introduced to the square lattice, i.e. $Q>0$, there exists a critical intensity of mobility when spiral waves do emerge, resulting in an oscillatory state which can be characterized by a nonzero $A>0$ value of the order parameter. Results presented in Fig.~\ref{random} highlight that for $Q=0.05$ global oscillations emerge when $\sigma$ becomes sufficiently large. Moreover, it can be observed that the impact of mobility can be so powerful that the system terminates into a homogeneous state ($A=1$) for high $\sigma$ values. Although at $Q=0.05$ (main panel) the small-world effect is already significant \cite{watts_dj_n98}, the inset shows that in fact just a tiny amount of randomness is enough to reach the oscillatory state if the mobility is sufficiently strong. More precisely, a nonzero value of $A$ can be detected already at $Q=0.002$.

\begin{figure}
\centerline{\epsfig{file=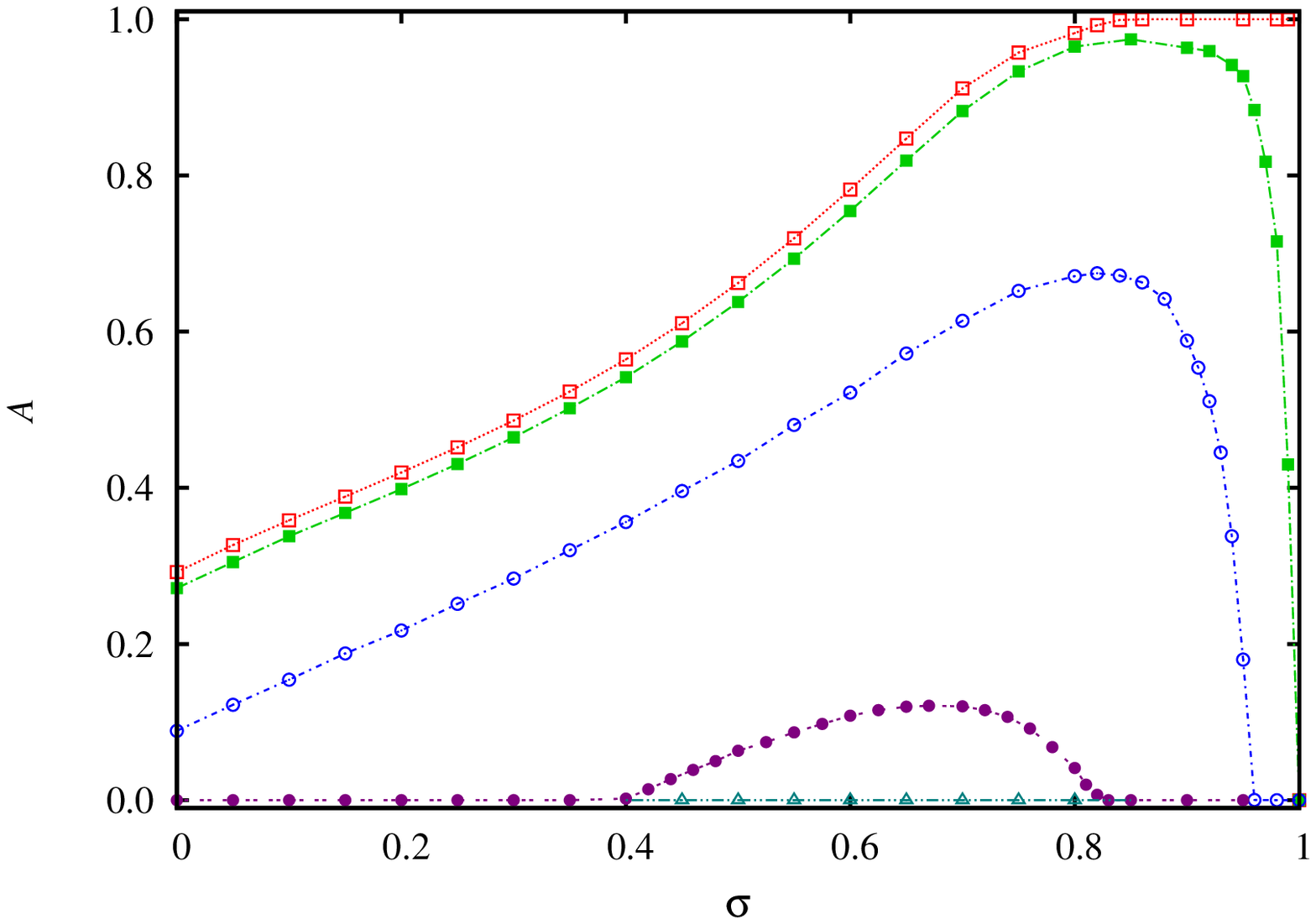,width=8.5cm}}
\centerline{\epsfig{file=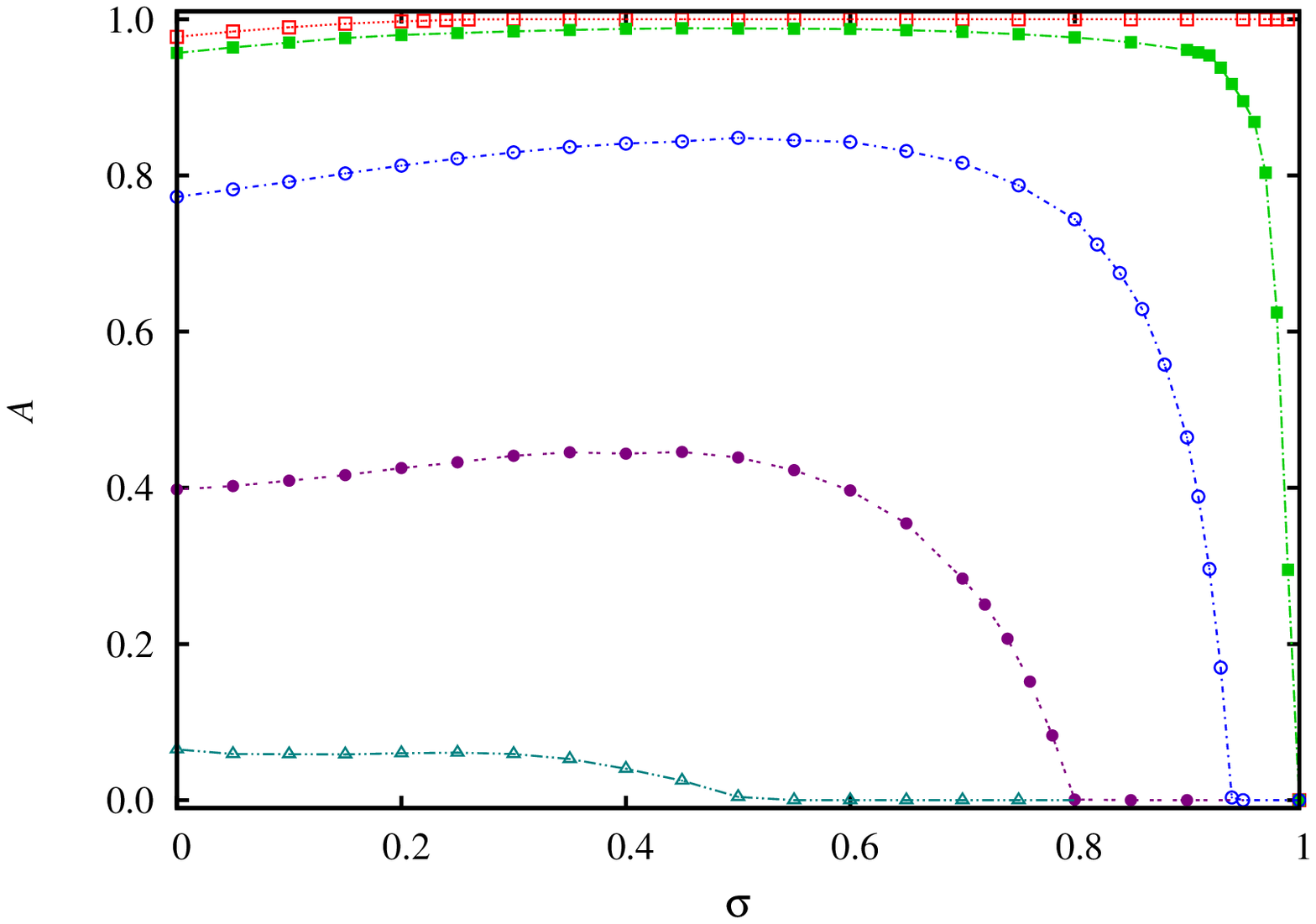,width=8.5cm}}
\caption{(Color online) Zealots effectively tame global oscillations, regardless of the fraction of rewired links and the intensity of mobility. Both panels show the area of the limit cycle in the ternary diagram $A$ in dependence on the intensity of mobility $\sigma$, as obtained for $Q=0.10$ (top) and $Q=0.99$ (bottom). The concentration of zealots in both panels is $\mu=0$, $0.001$, $0.01$, $0.03$, and $0.05$ from top to bottom.}
\label{mob}
\end{figure}

Naturally, the stronger the randomness in the interaction network, the more powerful the impact of mobility becomes. This is illustrated by the results presented in Fig.~\ref{mob} when the zealots are absent (upper-most curves in both panels, $\mu=0$). These results confirm that mobility indeed jeopardizes biodiversity, even if the total number of players is conserved, as long as there is a small amount of randomness present in the system to nucleate the spiral waves. As the wavelength of these waves approaches the linear size of the system, which we technically observe with $A \to 1$, termination into a homogeneous state becomes a likely evolutionary outcome.

The above conclusions might be somewhat uncomfortable because the diversity among competing strategies in a closed loop of dominance is a generally observed phenomenon. Accordingly, and since cyclical interactions were first theoretically raised precisely to bring about a potential source of diversity, we are motivated to find a mechanism to promote it. A viable option is to consider zealots \cite{mobilia_prl03, verma_pre15}. There are several reasons why such players who never change their strategy should be taken into account. Examples from human societies to microbial populations and excitable media are indeed plentiful. We thus designate a fraction $\mu$ of players who do not change strategy but serve only as potential sources of strategy invasion. Evidently, if the value of $\mu$ is too high then the system becomes trivial (nobody changes ever), so we restrict our study to small $\mu$ values.

Figure~\ref{mob} shows how drastically the stationary state changes when we introduce zealots. Firstly, the amplitude of oscillations always remains finite ($A$ stays below one), and this effect depends on the value of $\mu$ in a highly nonlinear manner. Even a tiny fraction of zealots (such as $0.001$) is capable to tame global oscillations efficiently, such that the amplitude of the order parameter $A$ is well below the $\mu=0$ reference curve at strong mobility. Secondly, it can be observed that the introduction of zealots always selects an optimal range of mobility, where oscillations with the largest amplitude can be observed. If we increase mobility further, then the oscillations become weaker, and the system gradually approaches the trivial $\sigma=1$ limit (there nothing happens, just the initial strategy distribution is mixed permanently). This effect is more pronounced for smaller values of $Q$ and becomes less visible at high $Q$ values, when the interaction network converges to the random regular graph. In view of the results presented in Fig.~\ref{mob}, a general conclusion with regards to the power of zealots is that a fraction of around 5\% is capable to completely tame oscillations in the system, no matter how random the interaction network or how intense the mobility.

\begin{figure*}
\centerline{\epsfig{file=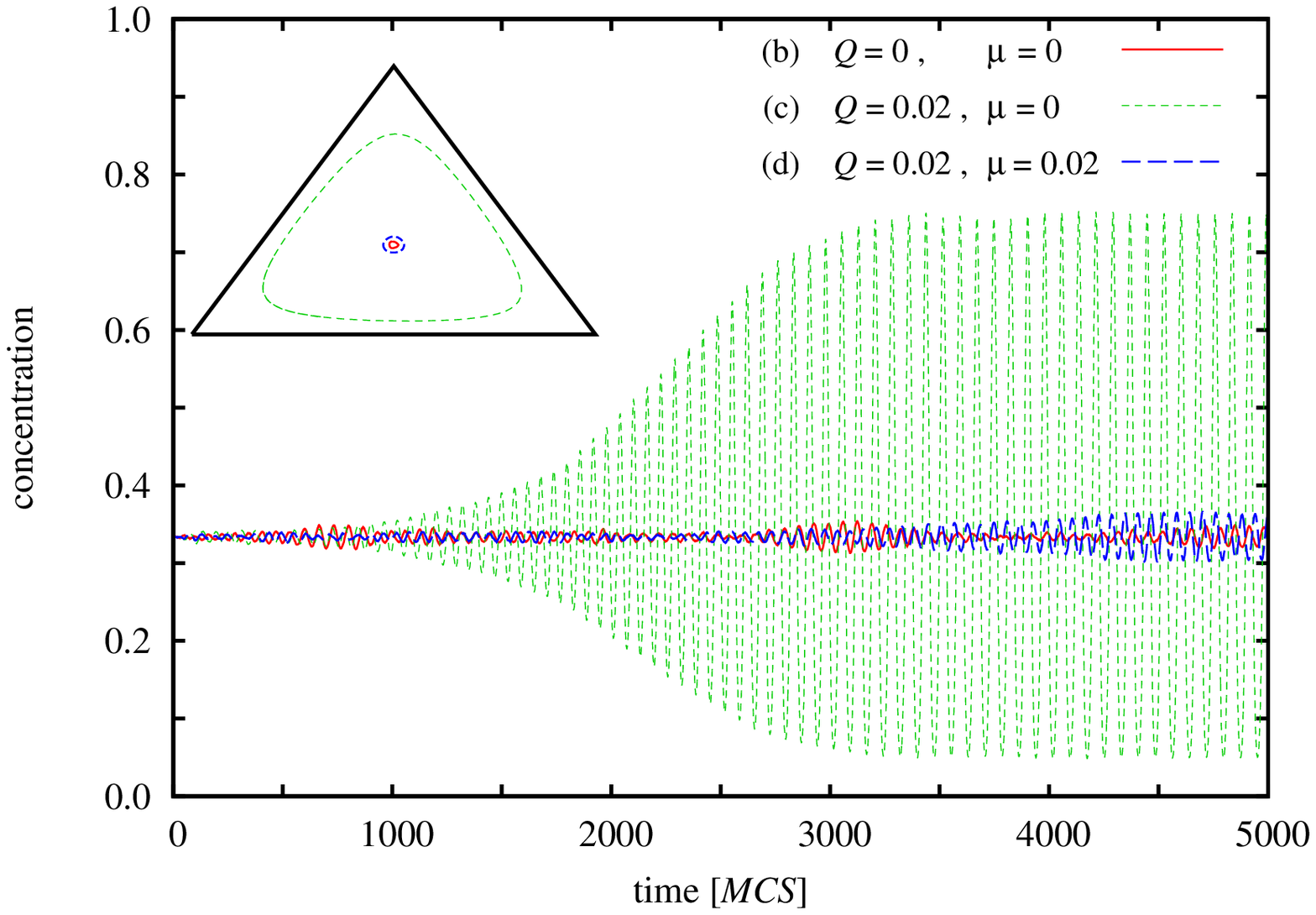,width=8.5cm}}
\centerline{\epsfig{file=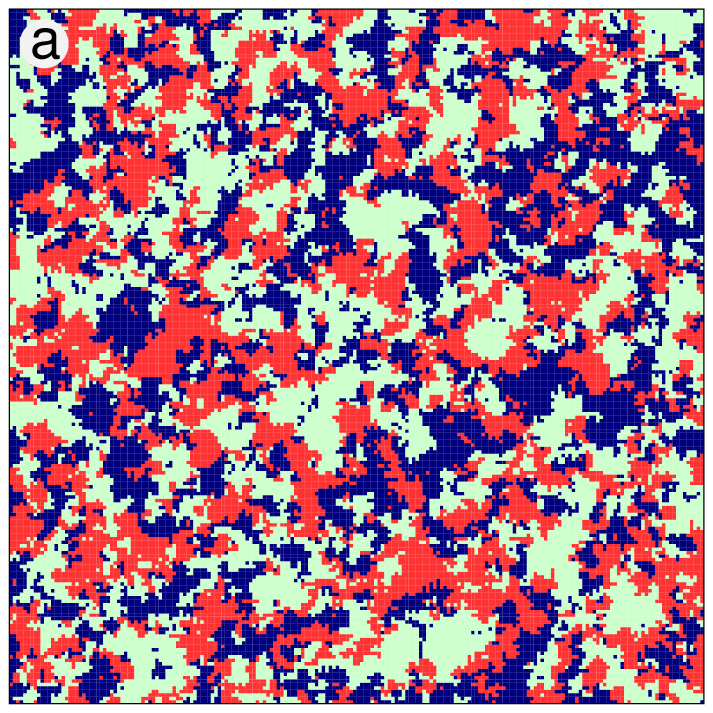,width=4.2cm}\epsfig{file=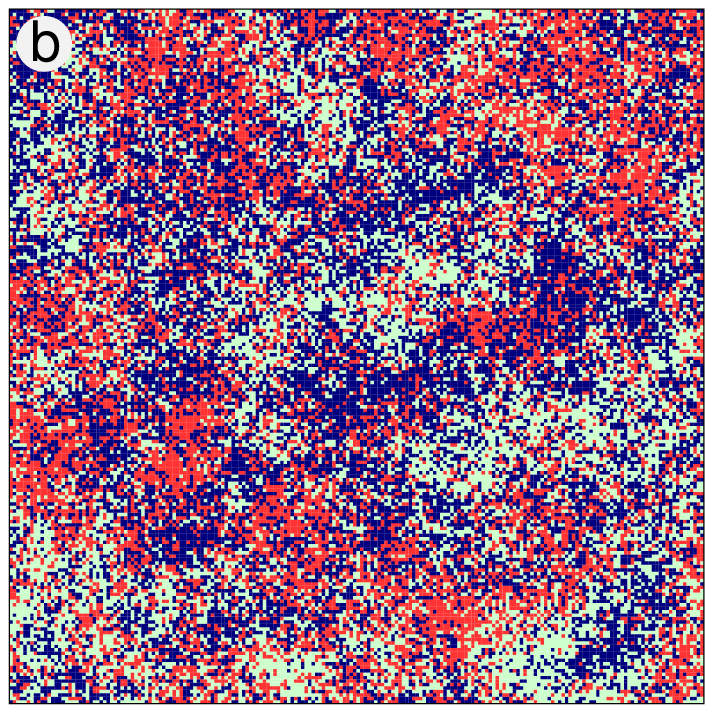,width=4.2cm}\epsfig{file=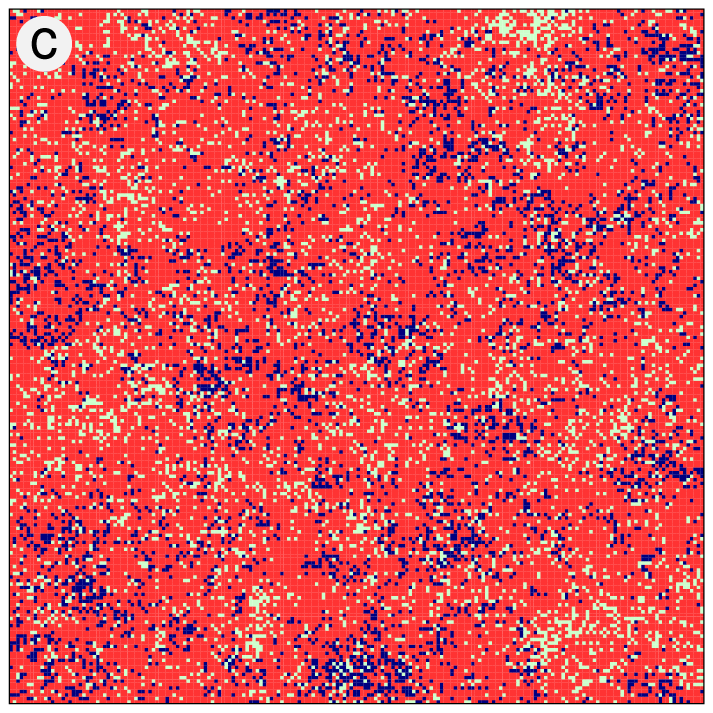,width=4.2cm}\epsfig{file=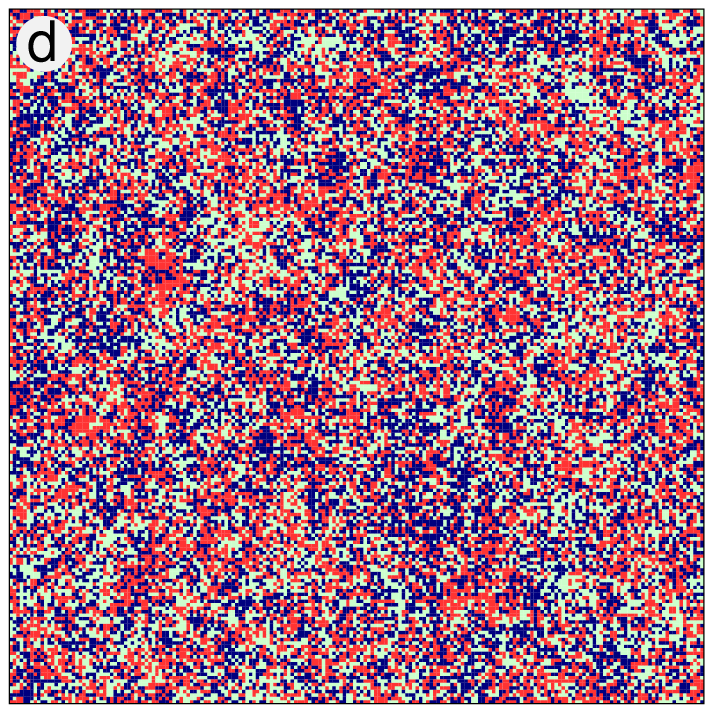,width=4.2cm}}
\caption{(Color online) Representative evolutionary outcomes in the spatial rock-paper-scissors game with zealots. The upper panel shows time courses of an arbitrary strategy (either $R$, $P$ or $S$), as obtained for different combinations of parameter values stated in the legend. The inset in the upper panel shows the corresponding trajectories in the ternary diagram, depicted with the same line style. In all three cases the intensity of mobility $\sigma=0.9$ was applied. The bottom row shows characteristic snapshots of the strategy distributions, corresponding to the cases indicated in the legend of the upper panel. Panel (a) in the bottom row shows the reference case for the classic spatial rock-paper-scissors game ($Q=0$, $\mu=0$, $\sigma=0$). For clarity, $200 \times 200$ excerpts of a larger $1000 \times 1000$ system are depicted in all four cases.}
\label{sum}
\end{figure*}

To better illustrate our findings, we present representative time courses of an arbitrary strategy in the upper panel of Fig.~\ref{sum}. In all three cases the evolution was launched from a random initial state. In the first case, marked by (b), the system remains around the central point of the ternary diagram, and this despite of the high intensity of mobility given by $\sigma=0.9$. Notably, the interaction network is a fully regular square lattice ($Q=0$), and since the number of players is conserved, this agrees with the results presented in \cite{peltomaki_pre08}. There it was pointed out first that the formation of spiral waves does not occur if there is a conservation law in place for the total number of competing players on a fully regular interaction network. In fact, by looking at the corresponding snapshot of strategy distributions depicted in panel (b) of the bottom row, we can see that the only consequence of intense mobility, if compared to the baseline case without mobility shown in panel (a) of the bottom row, is that the sharp interfaces separating competing domains evaporate. However, when a small amount of randomness is introduced to the regular lattice, then the system evolves towards a significantly different, global oscillatory state. This is case (c) in Fig.~\ref{sum}, where in addition to intense mobility also $Q=0.02>0$. We emphasize that the emergence of oscillations here is not an exclusive consequence of topological randomness, because in the absence of mobility the critical $Q=Q_c$ value for the oscillations to emerge is significantly higher, namely $Q_c=0.067$ \cite{szabo_jpa04}. Accordingly, there is synergy between topological randomness and high mobility that evokes the large amplitude oscillations. In the corresponding snapshot (c) in the bottom row, it is also illustrated nicely that one of the strategies (red in the present case) is temporarily dominant. Lastly, if just a small fraction of zealots is introduced while keeping all the other parameters the same, then the global oscillations vanish, and we arrive back to a state where diversity is not jeopardized anymore. This is illustrated by case (d) in Fig.~\ref{sum}, where $\mu=0.02$.

\begin{figure}
\centerline{\epsfig{file=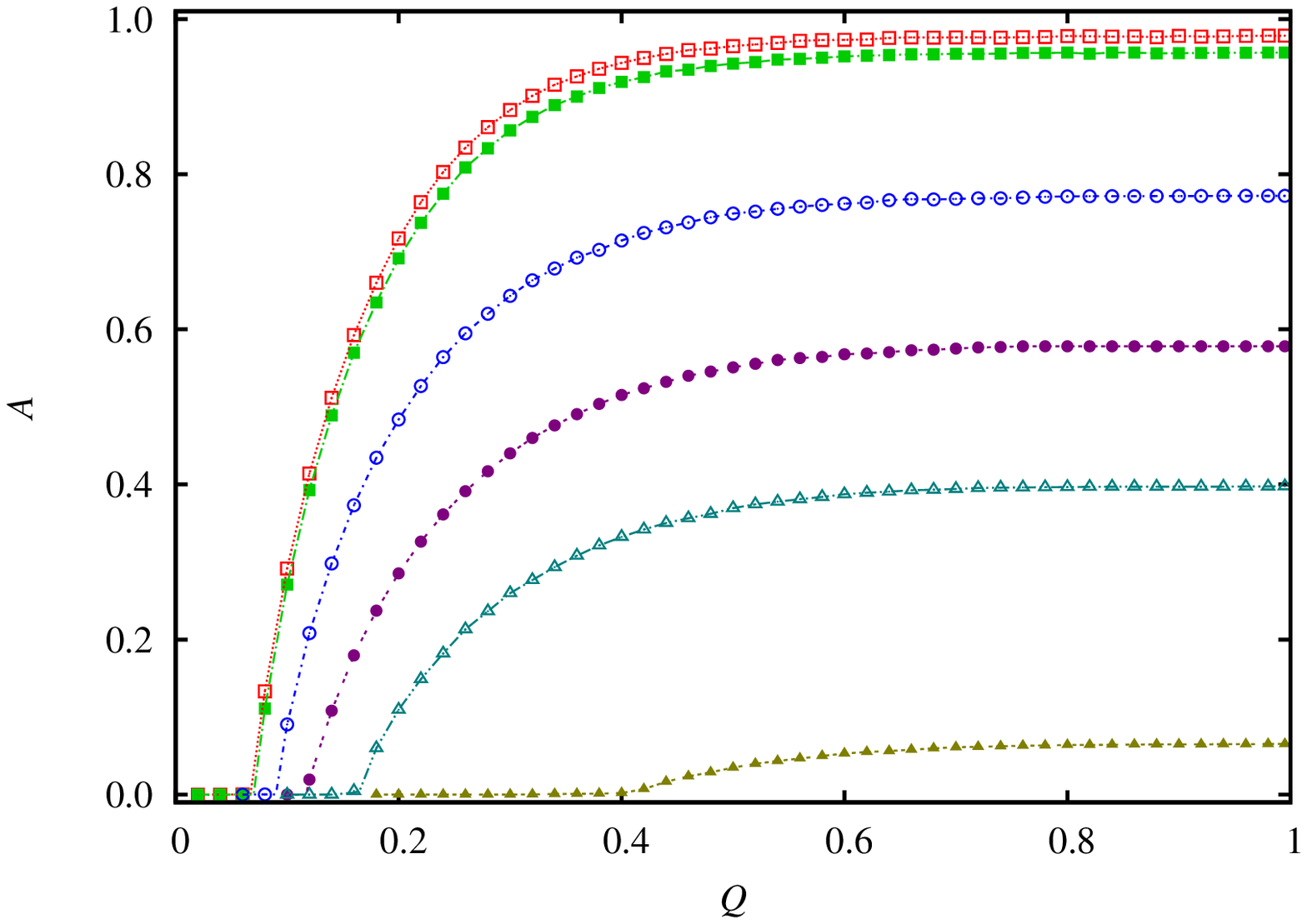,width=8.5cm}}
\centerline{\epsfig{file=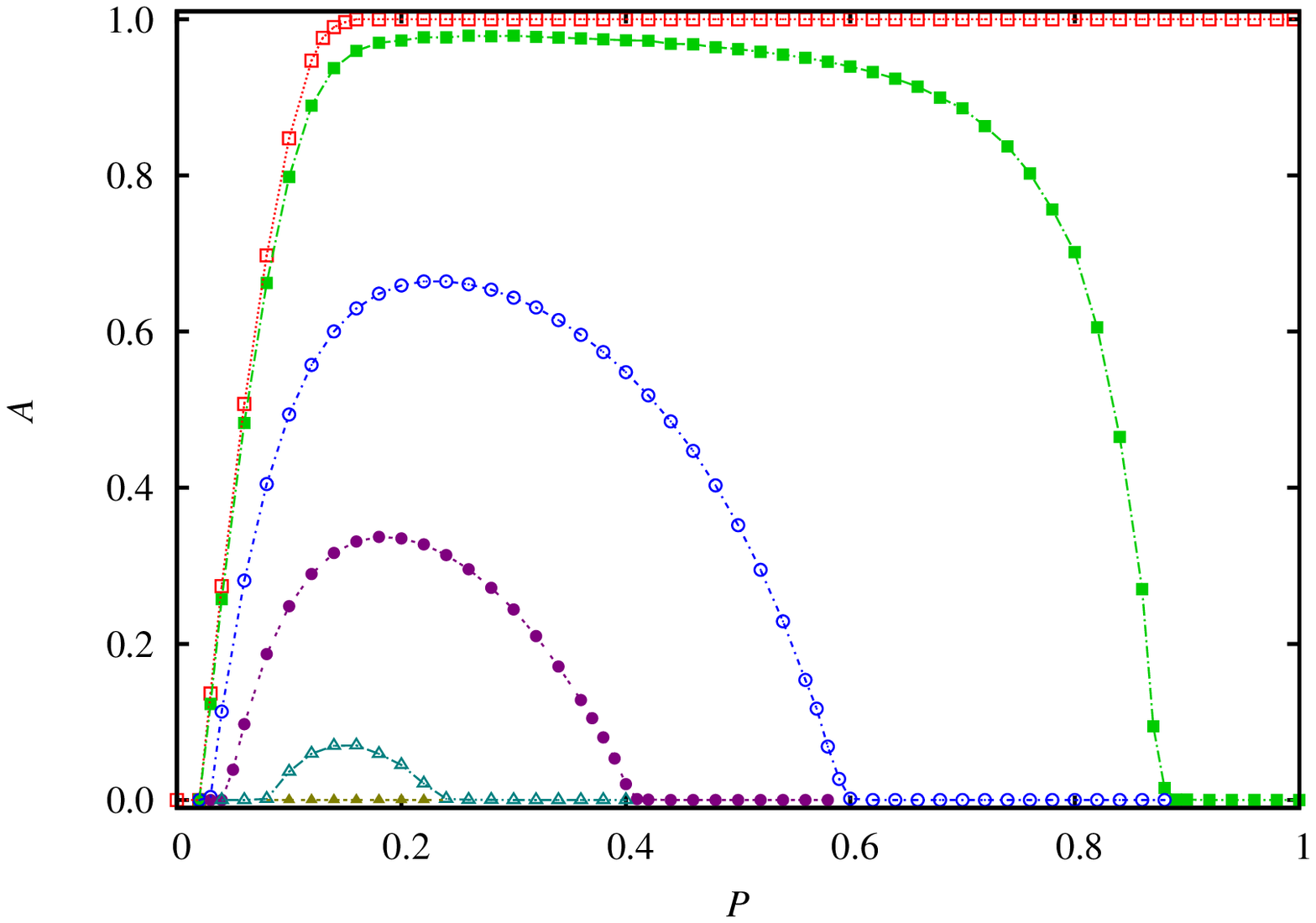,width=8.5cm}}
\caption{(Color online) The impact of zealots is qualitatively different when suppressed oscillations are due to quenched or annealed randomness. In the upper panel we show the area of the limit cycle in the ternary diagram $A$ in dependence on the fraction of rewired links $Q$ (quenched randomness), while in the lower panel we show $A$ in dependence on the probability $P$ that a target for an invasion is selected randomly from the whole population (annealed randomness). All presented results were obtained in the absence of mobility ($\sigma=0$), while the concentration of zealots in both panels is $\mu=0$, $0.001$, $0.01$, $0.02$, $0.03$ and $0.05$ from top to bottom.}
\label{randomness}
\end{figure}

We conclude our study by exploring how different types of topological randomness influence the impact of zealots in the stationary state. Since mobility is a source of randomness itself, we set $\sigma=0$ from here forth to have an unbiased comparison. By following previous studies \cite{zanette_pre01, szabo_jpa04, ohdaira_jpa14}, we consider quenched and annealed randomness, as described in Section~II. Results presented in Fig.~\ref{randomness} reveal that, in the absence of zealots ($\mu=0$), the central point of the ternary diagram becomes an unstable solution as we increase either type of topological randomness. It can be observed that as $Q$ and $P$ increase, a global oscillatory state becomes stable \cite{szabo_jpa04, szolnoki_pre04b}. But while the amplitude of oscillations always remains finite for quenched randomness (the order parameter $A$ always remains slightly smaller than one in the upper panel), the system always terminates into an absorbing homogeneous state above a critical $P$ value that determines annealed randomness ($A=1$ for sufficiently large $P$ values in the lower panel). As expected based on the results presented above, the introduction of zealots always successfully tames the oscillations, regardless of whether quenched or annealed randomness is applied. Already a minute $\mu=0.001$ fraction of zealots can preclude the system drifting off to a homogeneous single-strategy state, while for $\mu=0.05$ oscillations practically vanish altogether.

Nevertheless, there is a notable difference between the results presented in the upper and lower panel of Fig.~\ref{randomness}. In the upper panel, where quenched randomness is applied, the presence of zealots does not alter the behavior of the system qualitatively. In particular, larger values of $Q$ always increase the amplitude of oscillations, or at least the value of $A$ does not drop as $Q$ increases. This is not the case in the lower panel, where annealed randomness is applied. There, after introducing zealots, oscillations vanish and $A$ drops sharply when $P \to 1$. As we increase $\mu$, the intermediate $P$ interval where oscillations remain possible shrinks even further. Indeed, for annealed randomness the mean-field solution is always stable if the value of $P$ is sufficiently large. The shape of curves resembles what we have observed for mobility in Fig.~\ref{mob}. We therefore conclude that the impact of zealots in the presence of annealed randomness is qualitatively the same as when the mixing of strategies is due to mobility. The mutual feature of both, annealed randomness and mobility, is namely that fast diffusion does not necessarily destroy species coexistence. The presence of zealots thus helps to recover the original mean-field solution, thereby preserving diversity. Interestingly, even if the symmetry in zealotry is broken so that the three strategies are not equally represented among zealots, the limiting case being that only a single strategy contains zealots, our main results remain practically unchanged. This reveals that the main impact of zealots is blocking the propagation of waves, which in turn hinders global coordination to evolve. The strategy of zealots has thus only second-order importance in maintaining the diverse three-strategy state.

\section{Discussion}
We have studied the impact of mobility and zealotry in a spatial rock-paper-scissors game where the total number of players was conserved, and where the square lattice was in addition subject to quenched and annealed randomness. We have shown that the adverse impact of mobility is fully restored even if the total number of players is conserved, as long as the interaction lattice contains even a tiny amount of randomness. Detailed Monte Carlo simulations have revealed that, under strong mobility, as low as $2$ in $1000$ links are enough to be rewired for spiral waves to emerge. Naturally, we have also shown that the higher the interaction randomness, the lower the mobility needs to be for the same effect to emerge. In this sense, mobility and interaction randomness have the same effect, which is further corroborated by the fact that zealots have qualitatively the same impact when mitigating adverse effects of mobility, as they do when mitigating adverse effects of annealed randomness. Regardless of whether the promotion of spiral waves is due to quenched or annealed randomness, or due to mobility, our research reveals that zealots unambiguously suppress oscillations, thus contributing relevantly to the preservation of diversity. Interestingly, if only $1$ out of $1000$ players is a zealot, conditions already preclude extinction that would be due to large-amplitude system-wide oscillations. While even such a tiny fraction of zealots brings significant benefits, at 5\% occupancy zealots practically destroy all oscillations regardless of the intensity of mobility, and regardless of the type and strength of randomness in the interaction structure. Taken together, zealots are thus an important and highly effective asset for maintaining diversity in models of cyclic dominance.

Although many living systems can be adequately described solely by rock-paper-scissors-like intransitive relationships \cite{kerr_n02, kirkup_n04, neumann_g_dcdssb07, prado_evol08, cameron_jecol09, carranza_rsos16}, there also exist circumstances that require more realistic modeling. One option that was recently explored in the realm of cyclical interactions are the so-called protection spillovers \cite{kelsic_n15, bergstrom_n15, szolnoki_njp15}, which work under the assumption that rock can resist the invasion of paper if scissors are in the close neighborhood. Here we have further expanded the scope of possibilities by introducing zealots, which in non-living systems can be considered as local impurities that disobey otherwise valid laws, while in living systems they correspond to individuals with hardened, unchangeable types that are outside and unaffected by the closed loop of dominance. We expect that our findings will find relevance for patter formation in microbial populations \cite{durrett_jtb97, kirkup_n04, neumann_gf_bs10, nahum_pnas11}, in excitable media \cite{perc_njp07b}, and in human systems. Since the list of examples where the puzzle of biological diversity can be explained by cyclical interactions in the governing food webs is impressively long and inspiring \cite{stouffer_s12, szolnoki_jrsif14}, we hope that these recent theoretical explorations will inspire experimental work and further research along the same lines.

This research was supported by the Hungarian National Research Fund (Grant K-120785) and the Slovenian Research Agency (Grants J1-7009 and P5-0027).

\end{document}